\documentclass[aps, prd, superscriptaddress, twocolumn]{revtex4-1}

\usepackage{wallpaper}
\usepackage{bm}
\usepackage{amsmath}
\usepackage{amssymb}
\usepackage[colorlinks,linkcolor=blue,anchorcolor=blue,citecolor=red]{hyperref}

\usepackage{booktabs}
\usepackage{array} 
\usepackage{graphicx}

\begin{document}

\title{Multibaryon states in the framework of an equivparticle model}

\author{Hao-Song~You}
\affiliation{Center for Gravitation and Cosmology, College of Physical Science and Technology, Yangzhou University, Yangzhou 225009, China}
\affiliation{Tsung-Dao Lee Institute, Shanghai Jiao Tong University, Shanghai 201210, China}

\author{Xinmei Zhu}
\affiliation{College of Physical Science and Technology, Yangzhou University, Yangzhou 225009, China}

\author{Cheng-Jun~Xia}
\email{cjxia@yzu.edu.cn}
\affiliation{Center for Gravitation and Cosmology, College of Physical Science and Technology, Yangzhou University, Yangzhou 225009, China}

\author{Jialun Ping}
\affiliation{Department of Physics, Nanjing Normal University, Nanjing 210097, China}

\author{Ren-Xin Xu}
\affiliation{School of Physics and State Key Laboratory of Nuclear Physics and Technology, Peking University, Beijing 100871, China}
\affiliation{Kavli Institute for Astronomy and Astrophysics, Peking University, Beijing 100871, China}

\date{\today}

\begin{abstract}
Within the framework of an equivparticle model employing mean-field approximation, we investigate systematically the mass spectra of color-singlet $N$-quark configurations with $N = 3, 6, 9, 12, 15$, and 18, which are assumed to be spherically symmetric with quarks occupying the 1s$_{1/2}$ state, i.e., compact multibaryon states. At a given quark number $N$, these states collectively form a single irreducible representation under SU(6) symmetry. Our analysis yields comprehensive mass formulae that characterize these SU(6) multiplets, providing a unified description of their mass spectra. In order to effectively constrain the parameter space of the model and improve the prediction accuracy, we carry out a Bayesian parameter inference based on the experimental masses of eight baryons and $D_{03}$. The posterior
probability density functions and their correlations of the model parameters are examined, based on which we further predict the masses of various multibaryon states and provide their 68$\%$ and  90$\%$ credible intervals. In our prediction, H-dibaryon, $D_{03}$, and the dibaryon with $S = -6$ are all bound states relative to $\Lambda\Lambda$, $\Delta\Delta$ and $\Omega\Omega$ thresholds, while slight probabilities of other stable dibaryons (23.64$\%$ more stable than $\Omega^- \Xi^0$ for the state wtih $I = \frac{1}{2}$, $S = -5$ and 92.49$\%$ more stable than $\Xi^0\Xi^-$ for the state wtih $I = 0$, $S = -4$) and tribaryons (0.25$\%$ more stable than $\Xi^-\Xi^0\Xi^0$ for the state wtih $S = -6$, $I = 1/2$; 2.19$\%$ more stable than $\Lambda\Xi^0\Xi^-$ for the state wtih  $S = -5$, $I = 0$; and 2.21$\%$ more stable than $\Lambda\Lambda\Xi^0$ for the state wtih $S = -4$, $I = 1/2$) are observed as well. For heavier compact multibaryon states, it is unlikely for them to be stable. The stable and unstable multibaryon states examined in this work may persist in compact stars, which could be helpful for us to understand the essence of pulsar-like objects.
\end{abstract}


\maketitle

\section{\label{sec:intro}Introduction}
Quantum Chromodynamics (QCD) predicts the possible existence of exotic quark configurations beyond conventional mesons and baryons. These include multiquark states, multibaryons, quark nuggets, and even compact stellar objects composed of quark matter. However, due to the nonperturbative nature of QCD and the notorious sign problem in lattice simulations, our current understanding of such objects largely depends on various QCD-inspired effective models. Various theoretical approaches have been employed to study multibaryon systems, e.g., the bag model~\cite{Jaffe1977_PRL38-195,Jaffe1977_PRL38-617,Aerts1978_PRD17-260,Mulders1980_PRD21-2653,Liu1982_PLB113-1,MALTMAN1992_PLB291-371,Maezawa2005_PTP114-317}, nonrelativistic quark cluster model~\cite{OKA1983_PLB365-130,STRAUB1988_PLB200-241,OKA1991_NPA524-649,Shen1999_JPG25-1807}, Skyrme model~\cite{Balachandran1984_PRL52-887,JAFFE1985_NPB258-468,Yost1985_PRD32-816,KOPELIOVICH1992_NPA549-485}, and diquark model~\cite{Lee2009_EPJC64-283}. For strange multibaryons (strangeons), the interactions have been examined using diverse frameworks such as the quark cluster model~\cite{SAKAI1997_NPA625-192}, extended $\sigma-\omega-\rho$ mean-field model~\cite{Glendenning1998_PRC58-1298}, Lennard-Jones potential model~\cite{LAi2009_MNRAS398-L31},  linked bag model~\cite{Zhang1992_PRC46-2294,Miao2022_IJMPE31-2250037}, and so on. Theoretical investigations on the properties of quark matter have been conducted within several well-established frameworks, including the MIT bag model~\cite{Chodos1974_PRD10-2599,Farhi1984_PRD30-2379},  Nambu-Jona-Lasinio (NJL) model~\cite{Nambu1961_PR122-345},  perturbative QCD approaches~\cite{Fraga2013_APJL781-L25}, field correlator method~\cite{Plumari2013_PRD88-083005},  quasiparticle model~\cite{Goloviznin1992_ZPC57-671,Gorenstein1995_PRD52-5206,SCHERTLER1997_NPA616-659,Peshier2000_PRC61-045203,Bannur2007_PRC75-044905,GARDIM2009_NPA825-222},   equivparticle model~\cite{Fowler1981_ZPC9-271,Plumer1984_PLB3-198,CHAKRABARTY1989112,Benvenuto1995_PRD51-1989,Peng2000_PRC62-025801}, and so on.

Over the past few decades, various models have been used to predict the properties of multiquark states. For example, Dyson and Xuong extended the quark model to include 4-, 5-, and 6-quark states~\cite{Dyson1964_PRL13-815}. They simplified the Gursey--Radicati mass formula~\cite{Gursey_1964PRL13-173} to the SU(2) case and determined the spin- and isospin-dependent parameters by fitting the masses of the deuteron and the $N\Delta$ resonance $D_{12}$, subsequently predicting the masses of $D_{03}$ and $D_{30}$ to be around 2.35 GeV. In the SU(3) chiral constituent quark model with coupled $\Delta\Delta$ and $C_8C_8$ configurations, $d^*$ is predicted to be a compact hexaquark state with a binding energy of 47--84 MeV with its mass of 2380--2417 MeV, where the hidden-color component dominates (66--68$\%$)~\cite{Dong2013_PPNP_131-104045}. The pure $\Delta\Delta$ structure model yields smaller binding energies (29--62 MeV) and fails to explain the narrow width~\cite{Ping2009PRC79-024001}. The $D_{12}\pi$ molecular model proposed by Gal and Garcilazo (mass 2363$ \pm $20 MeV) requires mixing with a $\Delta\Delta$ core ($\alpha \sim 5/7$) to satisfy the experimental upper limit for single-pion decay branching ratios~\cite{Gal2013_PRL111-172301,Gal2014_NPA928-73}. Other approaches like lattice QCD (binding energy 30--40 MeV)~\cite{Gongyo2020_PLB811-135935} and QCD sum rules provide scattered results but generally support $d^*$ as a bound state below the $\Delta\Delta$ threshold~\cite{Bashkanov2009_PRL102-052301}, with its exact structure awaiting further experimental verification. The H-dibaryon was originally proposed as a stable state within the framework of the bag model~\cite{Jaffe1977_PRL38-195,Jaffe1977_PRL38-617}. However, recent lattice QCD simulations suggest that the H-dibaryon is either weakly bound or unbound~\cite{SASAKI2020_NPA998-121737,Green2021_PRL127-242003}. The dibaryon $\Omega\Omega$ with quantum numbers $S=-6$, $I=0$, and $J=0$ was predicted in the flavor SU(3) Skyrmion model~\cite{Kopeliovich1998_NPA639-75c} and later confirmed as a bound state in QDCSM calculations~\cite{Pang2002_PRC66-025201}. Recent (2+1)-flavor lattice QCD simulations with $m_\pi\simeq146$ MeV show this state exhibits overall attraction near the unitary regime, supporting its experimental search via pair-momentum correlations in heavy-ion collisions~\cite{Gongyo2018_PRL120-212001}. In a wide range of parameter spaces in models like bag model, it has been shown that strange quark matter (SQM), consisting of roughly equal numbers of $u$, $d$, and $s$ quarks, could be more stable than ordinary nuclear matter~\cite{Bodmer1971_PRD4-1601,Witten1984_PRD30-272,Terazawa1989_JPSJ58-3555,Dexheimer:2013eua}. This opens the possibility for the existence of exotic objects such as strangelets~\cite{Farhi1984_PRD30-2379,Berger1987_PRC35-213,Gilson1993_PRL71-332,Peng2006_PLB633-314,You2024_PRD109-034003}, nuclearites~\cite{Rujula1984_Nature312-5996,LOWDER1991_NPB24-177}, meteor-like compact ultradense objects~\cite{Rafelski2013_PRL110-111102}, and strange stars~\cite{Itoh1970_PTP44-291,Alcock1986_APJ310-261}.

The exotic objects mention above are expected to be formed under extreme conditions. The H-dibaryon may appear in high-density neutron star cores, requiring a balance between $\sigma$ attraction and $\omega$ repulsion, with its mass near the $\Lambda$-$\Lambda$ threshold but slightly above $26 \pm 11$ MeV~\cite{Inoue2012_NPA881-28,Beane2013_PRD87-034506,Shanahan2014_JPSCP1-013028}. RHIC's STAR measurements of $\Lambda\Lambda$ correlations in 200 GeV Au-Au collisions reveal slightly attractive interactions~\cite{Adamczyk2015_PRL114-022301,Morita2015_PRC91-024916}, while ALICE's dedicated H-dibaryon search (via $\mathrm{H}\rightarrow \Lambda p \pi^-$decay) in Pb-Pb collisions of LHC found no evidence, setting mass limits of 2200-2231 MeV (binding energy $<31$ MeV)~\cite{Adam2016_PLB752-267}. Zhang \textit{et al.} predicted a stable di-omega particle ($\Omega\Omega$) with strong binding ($\sim 100$ MeV) that could be formed in heavy-ion collisions, offering new possibilities to study exotic quark states~\cite{Zhang2000_PRC61-065204}. Vida\~{n}a \textit{et al.} examined the possible emergence of $d^*(2380)$ in neutron stars, predicting its presence in about 20$\%$ of heavy star cores, which reduces the mass of the most massive star by 15$\%$ and introduces new cooling mechanisms via neutrino emissions~\cite{Vidana2018_PLB781-112}. For SQM and nonstrange quark matter ($ud$QM), it may be formed through the following mechanisms in various high-energy processes, e.g., heavy-ion collisions~\cite{WEINER2006_IJMPE15-37}, mergers of binary compact stars~\cite{Madsen2002_JPG28-1737,Madsen2005_PRD71-014026,Bauswein2009_PRL103-011101,Lai2021_RAA21-250,Bucciantini2022_PRD106-103032}, Type II supernova explosions~\cite{Vucetich1998_PRD57-5959}, and the hadronization process in the early universe~\cite{Witten1984_PRD30-272}.

In this paper we consider the possible color-singlet $N$-quark states in the framework of an equivparticle model, where mean-field approximation (MFA) is employed~\cite{You2022_NPR39-302}. These $N$-quark states are assumed to be spherically symmetric with quarks occupying the 1s$_{1/2}$ state, which are thus all in positive parities. To fix the masses of baryons and multibaryon states, the correction from the color-magnetic part of one-gluon-exchange interaction needs to be  accounted for, where the coefficients for each flavor-spin multiplet should be fixed separately. As the allowed states must be totally antisymmetric with respect to flavor, spin, and color, the color-spin tensor operators in the mass formulae are expressed in simple flavor-spin tensor operators, where the contributions of the different SU(6)-breaking tensor operators can be identified. To constrain the model parameters, we further construct a likelihood function using the experimental masses of $8$ baryons and  $D_{03}$, and employ Bayesian inference to provide the posterior distribution of the parameter sets, where the mass spectra of multibaryon states can be estimated along with the corresponding posterior probability density functions.

This paper is organized as follows. In Sec.~\ref{sec:equivparticle}, we present the theoretical framework of the equivparticle model. The formulae to fix
the properties of baryons and multibaryon states are indicated in Sec.~\ref{Sec:MFA} with the center-of-mass corrections accounted for. Section~\ref{Sec:OGE} is devoted to the color-magnetic part of one-gluon-exchange interaction in the mass formulae. To fix the parameters and estimate the uncertainties of model predictions, in Sec.~\ref{sec:Bayesian} we illustrate the Bayesian inference approach adopted in this work. The obtained results on the properties baryons and multibaryon states are then presented in Sec.~\ref{sec:Results}. Finally, we give our conclusion in Sec.~\ref{sec:summary}.

\section{\label{Sec.theoretical} Theoretical framework}
\subsection{\label{sec:equivparticle}Equivparticle model}
\subsubsection{\label{sec:the_mass scaling} Quark mass scaling}
In the framework of the equivparticle model~\cite{Fowler1981_ZPC9-271,Plumer1984_PLB3-198,CHAKRABARTY1989112,Benvenuto1995_PRD51-1989,Peng2000_PRC62-025801,Peng1999_PRC61-015201,Wang2000_PRC62-015204,Zhang2001_EL56-361,LUGONES2003_IJMPD12-495,
Wen2005_PRD72-015204,Wen_2007,Yin2008_PRC77-055204,Chen2012_CPC36-947,Torres_2013,Chang:2013wnl,Xia2014_PRD89-105027,Chu2014_APJ780-135,Hou_2015,Peng:2016icu,Chu2017_PRD96-083019}, the strong interactions between quarks are described using density and/or temperature dependent quark masses.
In this approach, the mass of quark $i$ is determined by
\begin{equation}
\label{eq: mass scaling nb T}
m_{i} = m_{i0} + m_{\mathrm{I}}(\{ n_{j}, T \}).
\end{equation}
Here $n_j$ stands for the number density for quark flavor $j$, $T$ the temperature, and $m_{u0} = 3.415\  $MeV, $m_{d0} = 3.415\ $MeV, $m_{s0} = 93\ $MeV the current masses of quarks~\cite{Patrignani2016_CPC40-100001,PDG2020_PTEP2020-083C01}. To account for the effects of quark confinement, following the spirit of the bag model, the density-dependent quark masses for the zero-temperature case take the parametric form as
\begin{equation}
\label{eq: mass scaling bag}
m_{\mathrm{I}}(n_{\mathrm{b}}) = \frac{B}{3n_{\mathrm{b}}},
\end{equation}
where $B$ is the bag constant and the baryon number density $n_{\mathrm{b}} = \sum_{i = u,d,s} n_i/3$. Alternatively, a cubic-root scaling can be derived from considerations of linear confinement and the leading-order in-medium chiral condensate~\cite{Peng1999_PRC61-015201}, expressed as:
\begin{equation}
\label{eq: mass scaling cubic}
m_{\mathrm{I}}(n_{\mathrm{b}}) = D n_{\mathrm{b}}^{-1/3}.
\end{equation}
Here $D = -3(2/\pi)^{1/3}\sigma n^*/\sum_q\langle \bar{q} q\rangle_0$ represents the confinement strength with $\sigma$ being the string tension, $n^* = 0.49 \ $fm$^{-3}$ the chiral restoration density in the linear expression~\cite{Cohen1992_PRC45-1881}, and $\sum_q \langle \bar{q}q \rangle_0$ the sum of vacuum chiral condensates. Further consideration of one-gluon-exchange interaction suggests~\cite{Chen2012_CPC36-947}
\begin{equation}
\label{ea: mass scaling oge}
m_{\mathrm{I}}(n_{\mathrm{b}}) = D n_{\mathrm{b}}^{-1/3} - C n_{\mathrm{b}}^{1/3},
\end{equation}
where $C = -4(2/\pi)^{1/3}\alpha_s n^*/\sum_q\langle \bar{q}q \rangle_0$ varies with the strong coupling constant $\alpha_s$.  Incorporating the leading-order perturbative interactions at ultrahigh densities, we have found~\cite{Xia2014_PRD89-105027}
\begin{equation}
\label{eq: mass scaling perturbative}
m_{\mathrm{I}}(n_{\mathrm{b}}) = D n_{\mathrm{b}}^{-1/3} + C n_{\mathrm{b}}^{1/3},
\end{equation}
where $C \approx \pi^{2/3}\sqrt{{2\alpha_s}/{3\pi}}$. To investigate the influence of quark matter symmetry energy, an isospin-dependent term was incorporated, expressed as:
\begin{equation}
\label{ea: mass scaling isospin}
m_i(n_{\mathrm{b}},\delta) = m_{i0}+D n_{\mathrm{b}}^{-1/3}-\tau_i\delta D_I n_{\mathrm{b}}^{\alpha} \mathrm{e}^{-\beta n_{\mathrm{b}}},
\end{equation}
with $\tau_i$ being the third component of isospin for quark flavor $i$ and $\delta = 3(n_d-n_u)/(n_d+n_u)$ the isospin asymmetry~\cite{Chu2014_APJ780-135}. In our previous study, we have proposed a similar mass scaling~\cite{Wang2021_Galaxies9-70}, i.e.,
\begin{equation}
\label{eq: mass scaling isospin2}
m_{\mathrm{I}}(n_{\mathrm{b}},\delta) = D n_{\mathrm{b}}^{-1/3} +C n_{\mathrm{b}}^{1/3} + C_I \delta^2 n_{\mathrm{b}}.
\end{equation}
where $C_I$ corresponds to the strength of symmetry energy and is sensitive to the strong interactions among quarks~\cite{Cao2022_PRD106-083007,Yuan2022_PRD105-123004,Chu2014_APJ780-135,JEONG2016_NPA945-21,Chu2019_PRC99-035802,Wu2019_AIP2127_020032}.

\subsubsection{\label{sec:Lagrangian} Lagrangian density}
The lagrangian density of the equivparticle model is determined by
\begin{equation}
\label{eq: lagrangian density}
\mathcal{L} = \sum_i \bar{\psi}_i \left[ i\gamma^{\mu}\partial_{\mu}-m_i -eq_i \gamma^{\mu}A_{\mu} \right]\psi_i -\frac{1}{4}A_{\mu\nu}A^{\mu\nu},
\end{equation}
here $\psi_i$ represents the Dirac spinor of quark flavor $i$, $m_i$ the mass, $q_i$ the charge ($q_u = 2e/3$ and $q_d = q_s = -e/3$), and $A_{\mu}$ the photon field with the field tensor
\begin{equation}
\label{eq: photon field}
A_{\mu\nu} = \partial_{\mu}A_{\nu}-\partial_{\nu}A_{\mu}.
\end{equation}

The equivparticle model accounts for strong interactions by introducing density-dependent quark masses, where the mass term in Eq. (\ref{eq: lagrangian density}) varies as a function of baryon number density:
\begin{equation}
\label{eq: mass scaling final}
m_i(n_{\mathrm{b}}) = f m_{i0} + D n_{\mathrm{b}}^{-1/3} + C n_{\mathrm{b}}^{1/3}.
\end{equation}
Note that we employ Eq. (\ref{eq: mass scaling perturbative}) for the mass scaling, which includes linear confinement and leading-order perturbative interactions. To accommodate running mass effects, the current quark masses are multiplied by a factor $f$~\cite{Politzer1973_PRL30-1346,Weinberg1973_PRD8-3497}.

Adopting the Euler-Lagrange equation, one obtains the equation of motion for quarks
\begin{equation}
\label{eq: motion-quark}
\gamma^{\mu} (i\partial_{\mu}-eq_iA_{\mu}-V_V\delta_{\mu0} )\psi_i = m_i \psi_i.
\end{equation}
The term $V_V$ arises from the density dependent quark masses, i.e,
\begin{equation}
\label{eq: V_v-psi}
V_V = \frac{1}{3}\frac{\mathrm{d}m_{\mathrm{I}}}{\mathrm{d}n_{\mathrm{b}}}\sum_i \bar{\psi}_i\psi_i.
\end{equation}
The equation of motion for photons is given by
\begin{equation}
\label{eq: motion-photon}
\square A^{\mu} = e\sum_i q_i\bar{\psi}_i\gamma^{\mu}\psi_i = e\sum_i J_i^{\mu e}
\end{equation}

\subsection{\label{Sec:MFA} Baryon and multibaryon states}
\subsubsection{\label{sec:symmetric case}Spherically symmetric systems in MFA}
For spherically symmetric systems, the Dirac spinor of quarks can be expanded as
\begin{equation}
 \psi_{n\kappa m}({\bm r}) =\frac{1}{r}
 \left(\begin{array}{c}
   iG_{n\kappa}(r) \\
    F_{n\kappa}(r) {\bm\sigma}\cdot{\hat{\bm r}} \\
 \end{array}\right) Y_{jm}^l(\theta,\phi)\:,
\label{EQ:RMF}
\end{equation}
where $G_{n\kappa}(r)/r$ and $F_{n\kappa}(r)/r$ are the radial wave functions for upper and lower components, and $Y_{jm}^l(\theta,\phi)$ is the spinor spherical harmonics, i.e.,
\begin{equation}
\label{eq: spherical harmonics}
Y_{jm}^l(\theta,\phi) = \sum_{l_z,s_z}\langle l,l_z;1/2,s_z|j,m \rangle Y_{ll}\chi_{1/2}^{s_z}
\end{equation}
The quantum number $\kappa$ is defined by the angular momenta ($l,j$) as $\kappa = (-1)^{j+l+1/2}(j+1/2)$ with $j=l\pm1/2$. 
By utilizing MFA and inserting Eq.~(\ref{EQ:RMF}) into the Dirac equation, we can easily obtain the one-dimensional radial Dirac equation through the integration of the angular component, i.e.,
\begin{equation}
 \left(\begin{array}{cc}
  V_{iV} + V_{iS}                                                   & {\displaystyle -\frac{\mbox{d}}{\mbox{d}r} + \frac{\kappa}{r}}\\
  {\displaystyle \frac{\mbox{d}}{\mbox{d}r}+\frac{\kappa}{r}} & V_{iV} - V_{iS}                       \\
 \end{array}\right)
 \left(\begin{array}{c}
  G_{in\kappa} \\
  F_{in\kappa} \\
 \end{array}\right)
 = \varepsilon_{in\kappa}
 \left(\begin{array}{c}
  G_{in\kappa} \\
  F_{in\kappa} \\
 \end{array}\right) \:,
\label{Eq:RDirac}
\end{equation}
where $\varepsilon_{in\kappa}$ is the single particle energy. The mean-field scalar and vector potentials of quarks can be obtained as
\begin{eqnarray}
 V_{iS} &=& f m_{i0} +
 m_\mathrm{I}(n_\mathrm{b}), \label{Eq:Vs}\\
 V_{iV} &=& \frac{1}{3}\frac{\mbox{d} m_\mathrm{I}}{\mbox{d} n_\mathrm{b}}\sum_{i=u,d,s}  n_i^\mathrm{s} + e q_i A_0, \label{Eq:Vv}
\end{eqnarray}
where there exist common terms in the scalar and vector potentials, i.e., $V_{S} = m_\mathrm{I}(n_\mathrm{b})$ and $V_{V} = \frac{1}{3}\frac{\mbox{d} m_\mathrm{I}}{\mbox{d} n_\mathrm{b}}\sum_{i=u,d,s}  n_i^\mathrm{s}$.
In addition, it should be noted that the scalar potential in Eq.~(\ref{Eq:Vs}) now incorporates the current masses of quarks. The vector potential arises as a result of the density-dependence of quark masses and is crucial for ensuring thermodynamic self-consistency~\cite{Xia2018_PRD98-034031,Xia2014_PRD89-105027}. The Klein-Gordon equation for photons is given by
\begin{equation}
- \nabla^2 A_0 = e n_\mathrm{ch}. \label{Eq:K-G}
\end{equation}
where $n_\mathrm{ch}=\sum_iq_in_i$ is the charge density.

We then adopt no-sea approximation and disregard any contributions from anti-quarks. For given radial wave functions, the scalar and vector densities are fixed by
\begin{subequations}
\begin{eqnarray}
 n_i^\mathrm{s}(r) &=& \sum_{n,\kappa} \frac{3|\kappa|}{2\pi r^2}
 \left[|G_{in\kappa}(r)|^2-|F_{in\kappa}(r)|^2\right] \:,
\\
 n_i(r) &=& \sum_{n,\kappa} \frac{3|\kappa|}{2\pi r^2}
 \left[|G_{in\kappa}(r)|^2+|F_{in\kappa}(r)|^2\right] \:,
\end{eqnarray}%
\label{Eq:Density}%
\end{subequations}%
where the degeneracy factor of each single particle levels is $3(2j+1)=6|\kappa|$. The quark numbers $N_i=\int 4\pi r^2 n_i(r)dr$ $(i=u,d,s)$ are obtained by integrating the density $n_i(r)$ in coordinate space.

Finally, the total mass and free energy in MFA can be obtained with
\begin{eqnarray}
E_0 &=& \sum_{i,n,\kappa} 6|\kappa|\varepsilon_{in\kappa} - \int 12\pi r^2 n_\mathrm{b}(r) V_V(r) \mbox{d}r \label{Eq:M} \\
  &\mathrm{}& - \int 2\pi r^2 n_\mathrm{ch}(r) e A_0(r) \mbox{d}r, \nonumber
\end{eqnarray}
where the last term represents the contribution of Coulomb energy. At fixed  parameters $C$, $D$, and $f$, we numerically solve the coupled system comprising the Dirac Eq. (\ref{Eq:RDirac}), mean-field potential Eqs. (\ref{Eq:Vs}), (\ref{Eq:Vv}), and density Eqs. (\ref{Eq:Density}) within a finite box by iteration in coordinate space with the grid width 0.001 fm.

\subsubsection{\label{sec:c-mass-c} Center-of-mass correction}
The obtained solution for a color-singlet $N$-quark state can be viewed as a wave packet of the physical particle with various total momentum $\vec{P}$, where the actual mass $M_0$ can be approximately obtained with~\cite{Bartelske1984_PRD29-1035}
\begin{equation}
\label{ea:E-mass}
E_0 = \left\langle \sqrt{M_0^2+P^2} \right\rangle \approx \sqrt{M_0^2+\langle P^2 \rangle},
\end{equation}
where $E_0$ is determined by Eq. (\ref{Eq:M}). The expectation value of the squared total momentum $\langle P^2 \rangle$ is obtained as described in Ref.~\cite{Bender2000}. Subsequently, by introducing $\Gamma$ with
\begin{equation}
\label{eq:gamma-corr}
\frac{1}{\Gamma} = \left\langle \frac{1}{\gamma} \right\rangle = \langle \sqrt{1-\nu^2} \rangle \approx \sqrt{1-\frac{\langle P^2 \rangle}{E_0^2}},
\end{equation}
the mean-square radius and mass with  center-of-mass corrections are obtained with
\begin{equation}
\label{eq:R2-corr}
\langle r^2 \rangle = \left[ \langle r^2 \rangle_0 - \frac{9Q}{4M_0^2(\Gamma^2-1)} \right]\frac{3\Gamma^2}{2\Gamma^2+1},
\end{equation}
\begin{equation}
\label{eq:actual-M}
M_0 = \frac{E_0}{\Gamma},
\end{equation}
where $M_0$ stands for the mass of the corresponding baryon or multibaryon state and $Q$ the charge number.

\subsection{\label{Sec:OGE} One-gluon-exchange interactions}
\subsubsection{\label{sec:quark classification} Classification of  $N$-quark states}
Here we consider the light quarks ($u,d,s$) occupying only the $1s_{1/2}$ orbital, which can accommodate a maximum of 18 quarks. Therefore,  the $N$-quark states can be decomposed into flavor ($F$), spin ($J$), and color ($C$) classifications as follows~\cite{Aerts1978_PRD17-260}: \begin{equation}
\label{eq:SU18}
\mathrm{SU}(18) \supset \mathrm{SU}(3, F) \otimes \mathrm{SU}(2, J) \otimes \mathrm{SU}(3, C),
\end{equation}
To take into account certain quantum numbers of these states, we consider decomposing SU(18) as follows:
\begin{equation}
\label{eq:SU18-2}
\mathrm{SU}(18) \supset \mathrm{SU}(6,FJ) \otimes \mathrm{SU}(3,C).
\end{equation}
Since we are dealing with SU(3,$C$) singlet states and quarks must be
totally antisymmetric with respect to flavor, spin, and color, the irreducible representation (irrep) of SU(6,$FJ$) can be fixed uniquely. We further decompose SU(6,$FJ$) into
\begin{equation}
\label{eq:SU6}
\mathrm{SU}(6,FJ) \supset \mathrm{SU}(3,F) \otimes \mathrm{SU}(2,J).
\end{equation}
The decomposition for $\mathrm{SU}(6,FJ)$ irreps in flavor and spin is then presented in Table~\ref{tab:su(3)-su(2)}.

\begin{table}[ht]
\centering
\footnotesize
\caption{ The decomposition of SU(6,$FJ$) irreps in flavor and spin at fixed baryon numbers $B=N/3$. }
\label{tab:su(3)-su(2)}
\begin{tabular}{@{}c >{\raggedright\arraybackslash}p{0.95\linewidth}@{}}
\hline
\hline
\multicolumn{1}{c}{$B$} & \multicolumn{1}{c}{Decomposition}\\ \hline
1 &
  $\begin{aligned}[t]
  [3]  =& [21] \otimes [21] + [3] \otimes [3] \\
  56  =& 8 \otimes 2 + 10 \otimes 4
  \end{aligned}$ \\[2ex]
\hline
2 &
  $\begin{aligned}[t]
  [33]  =& [6] \otimes [33] + [51] \otimes [42] + [42] \otimes [51] + [42] \otimes [33] \\
  &+ [33] \otimes [6] + [33] \otimes [42] + [411] \otimes [42] + [321] \otimes [51] \\
  &+ [321] \otimes [42] + [222] \otimes [33] \\
  490  =& 28 \otimes 1 + 35 \otimes 3 + 27 \otimes 5 + 27 \otimes 1 + 10^{*} \otimes 7 \\
  &+ 10^{*} \otimes 3 + 10 \otimes 3 + 8 \otimes 5 + 8 \otimes 3 + 1 \otimes 1
  \end{aligned}$ \\[2ex]
\hline
3 &
  $\begin{aligned}[t]
  [333]  =& [333] \otimes [9] + [333] \otimes [72] + [333] \otimes [63] + [432] \otimes [81] \\
  &+ [432] \otimes [72] + [432] \otimes [63] + [432] \otimes [54] + [522] \otimes [63] \\
  &+ [441] \otimes [63] + [531] \otimes [72] + [531] \otimes [63] + [531] \otimes [54] \\
  &+ [621] \otimes [54] + [54] \otimes [54] + [63] \otimes [63] \\
  980  =& 1 \otimes 10 + 1 \otimes 6 + 1 \otimes 4 + 8 \otimes 8 + 8 \otimes 6 + 8 \otimes 4 \\
  &+ 8 \otimes 2 + 10 \otimes 4 + 10^{*} \otimes 4 + 27 \otimes 6 + 27 \otimes 4 \\
  &+ 27 \otimes 2 + 35 \otimes 2 + 35^{*} \otimes 2 + 64 \otimes 4
  \end{aligned}$ \\[2ex]
\hline
4 &
  $\begin{aligned}[t]
  [3333]  =& [66] \otimes [66] + [651] \otimes [75] + [642] \otimes [84] + [642] \otimes [66] \\
  &+ [633] \otimes [93] + [633] \otimes [75] + [552] \otimes [75] + [543] \otimes [84] \\
  &+ [543] \otimes [75] + [444] \otimes [66] \\
  490  =& 28^{*} \otimes 1 + 35^{*} \otimes 3 + 27 \otimes 5 + 27 \otimes 1 + 10^{*} \otimes 7 \\
  &+ 10^{*} \otimes 3 + 10 \otimes 3 + 8 \otimes 5 + 8 \otimes 3 + 1 \otimes 1
  \end{aligned}$ \\[2ex]
\hline
5 &
  $\begin{aligned}[t]
  [33333]  =& [654] \otimes [87] + [663] \otimes [96] \\
  56^{*}  =& 8 \otimes 2 + 10^{*} \otimes 4
  \end{aligned}$ \\[2ex]
\hline
6 &
  $\begin{aligned}[t]
  [333333]  =& [666] \otimes [99] \\
  1  =& 1 \otimes 1
  \end{aligned}$ \\
\hline
\hline
\end{tabular}
\end{table}

\begin{table}[ht]
\centering
\footnotesize
\caption{The hypercharge, SU(4,$J_n$), and strange quark spin content of the SU(6,$FJ$) irreps. }
\label{tab:Y-SU(4)-Js)}
\begin{tabular}{@{}c >{\raggedright\arraybackslash}p{0.95\linewidth}@{}}
\hline
\hline
\multicolumn{1}{c}{$B$} & \multicolumn{1}{c}{Decomposition} \\
\hline
1 &
  $\begin{aligned}[t]
  [3]  =& (1)\otimes[3]\otimes[0] + (0)\otimes[2]\otimes[1] \\
  &+ (-1)\otimes[1]\otimes[2] + (-2)\otimes[0]\otimes[3] \\
  56  =& 20\otimes 1 + 10\otimes 2 + 4\otimes 3 + 1\otimes 4
  \end{aligned}$ \\
\hline
2 &
  $\begin{aligned}[t]
  [33]  =& (2)\otimes[33]\otimes[0] + (1)\otimes[32]\otimes[1] + (0)\otimes[31]\otimes[2]\\
  &+ (0)\otimes[22]\otimes[11] + (-1)\otimes[3]\otimes[3] + (-1)\otimes[21]\otimes[21]\\
  &+ (-2)\otimes[2]\otimes[31] + (-2)\otimes[11]\otimes[22] + (-3)\otimes[1]\otimes[32] \\
  &+ (-4)\otimes[0]\otimes[33] \\
  490  =& 50\otimes 1 + 60\otimes 2 + 45\otimes 3 + 20_2\otimes 1 + 20\otimes 4 \\
  &+ 20_1\otimes 2 + 10\otimes 3 + 6\otimes 1 + 4\otimes 2 + 1\otimes 1
  \end{aligned}$ \\
\hline
3 &
  $\begin{aligned}[t]
  [333]  =& (3)\otimes[333]\otimes[0] + (2)\otimes[332]\otimes[1] + (1)\otimes[331]\otimes[2] \\
  &+ (1)\otimes[322]\otimes[11] + (0)\otimes[321]\otimes[21] + (0)\otimes[33]\otimes[3] \\
  &+ (-1)\otimes[311]\otimes[22] + (-1)\otimes[32]\otimes[31] \\
  &+ (-2)\otimes[31]\otimes[32] + (-3)\otimes[3]\otimes[33] \\
  980  =& 20^*\otimes 1 + 45^*\otimes 2 + 60^*\otimes 3 + 36^*\otimes 1 + 64\otimes 2 \\
  &+ 50\otimes 4 + 36\otimes 1 + 60\otimes 3 + 45\otimes 2 + 20\otimes 1
  \end{aligned}$ \\
\hline
4 &
  $\begin{aligned}[t]
  [3333]  =& (4)\otimes[3333]\otimes[0] + (3)\otimes[3332]\otimes[1] \\
  &+ (2)\otimes[3331]\otimes[2] + (2)\otimes[3322]\otimes[11] \\
  &+ (1)\otimes[333]\otimes[3] + (1)\otimes[3321]\otimes[21] \\
  &+ (0)\otimes[332]\otimes[31] + (0)\otimes[3311]\otimes[22] \\
  &+ (-1)\otimes[331]\otimes[32] + (-2)\otimes[33]\otimes[33] \\
  490  =& 1^*\otimes 1 + 4^*\otimes 2 + 10^*\otimes 3 + 6^*\otimes 1 + 20^*\otimes 4 \\
  &+ 20^*\otimes 2 + 45^*\otimes 3 + 20^*\otimes 1 + 60^*\otimes 2 + 50\otimes 1
  \end{aligned}$ \\
\hline
5 &
  $\begin{aligned}[t]
  [33333]  =& (2)\otimes[3333]\otimes[3] + (1)\otimes[3332]\otimes[31] \\
  &+ (0)\otimes[3331]\otimes[32] + (-1)\otimes[333]\otimes[33] \\
  56^*  =& 1^*\otimes 4 + 4^*\otimes 3 + 10^*\otimes 2 + 20^*\otimes 1
  \end{aligned}$ \\
\hline
6 &
  $\begin{aligned}[t]
  [333333]  =& (0)\otimes[3333]\otimes[33] \\
  1^*  =& 1^*\otimes 1
  \end{aligned}$ \\
\hline
\hline
\end{tabular}
\end{table}

Consider the fact that SU(3,$F$) symmetry is broken, we adopt another decomposition, i.e.,
\begin{equation}
    \label{eq:Su(6) version2}
    \mathrm{SU}(6,FJ) \supset U(1,Y)\otimes \mathrm{SU}(4,IJ_n)\otimes \mathrm{SU}(2,J_s),
\end{equation}
where $Y$ is the hypercharge, $I$ the isospin, $J_n$ the total spin of nonstrange quarks, and $J_s$ the total spin of strange quarks. The decomposition for $\mathrm{SU}(6,FJ)$ irreps 
\begin{equation}
[\mu] = \sum_\oplus (Y,(\nu),J_s)
\end{equation}
is then indicated in Table~\ref{tab:Y-SU(4)-Js)}, where $(\nu)$ corresponds to SU(4,$IJ_n$) irreps. We further decompose SU(4,$IJ_n$) according to the isospin symmetry and the non-strange quark spin symmetry, i.e.,
\begin{equation}
    \label{eq:SU(4)}
    \mathrm{SU}(4,IJ_n) \supset \mathrm{SU}(2,I)\otimes \mathrm{SU}(2,J_n).
\end{equation}
These decompositions are given in Table~\ref{tab:su4-decompositions}.

\begin{table}[ht]
\centering
\footnotesize
\renewcommand{\arraystretch}{1.5} 
\setlength{\defaultaddspace}{2pt} 
\caption{The decomposition of SU(4,$IJ_{n}$) irreps in isospin and nonstrange spin.}
\label{tab:su4-decompositions}
\begin{tabular}{lll}
\hline
\hline
\multicolumn{1}{c}{Irrep} & \multicolumn{1}{c}{Dim} & \multicolumn{1}{c}{$(I,J_n)$}\\
\hline
$[0]$ & $(1)$ & $(0,0)$ \\
$[1]$ & $(4)$ & $(\frac{1}{2},\frac{1}{2})$ \\
$[2]$ & $(10)$ & $(1,1)\oplus(0,0)$ \\
$[3]$ & $(20)$ & $(\frac{3}{2},\frac{3}{2})\oplus(\frac{1}{2},\frac{1}{2})$ \\
$[11]$ & $(6)$ & $(1,0)\oplus(0,1)$ \\
$[21]$ & $(20_1)$ & $(\frac{3}{2},\frac{1}{2})\oplus(\frac{1}{2},\frac{3}{2})\oplus(\frac{1}{2},\frac{1}{2})$ \\
$[22]$ & $(20_2)$ & $(2,0)\oplus(1,1)\oplus(0,2)\oplus(0,0)$ \\
$[31]$ & $(45)$ & $(2,1)\oplus(1,2)\oplus(1,1)\oplus(1,0)\oplus(0,1)$ \\
$[32]$ & $(60)$ & $(\frac{5}{2},\frac{1}{2})\oplus(\frac{3}{2},\frac{3}{2})\oplus(\frac{3}{2},\frac{1}{2})\oplus(\frac{1}{2},\frac{5}{2})\oplus(\frac{1}{2},\frac{3}{2})\oplus(\frac{1}{2},\frac{1}{2})$ \\
$[33]$ & $(64)$ & $(2,1)\oplus(2,0)\oplus(1,2)\oplus 2(1,1)\oplus(1,0)\oplus(0,2)\oplus(0,1)$ \\
$[111]$ & $(10)$ & $(1,1)\oplus(0,0)$ \\
$[211]$ & $(20_2)$ & $(2,0)\oplus(1,1)\oplus(0,2)\oplus(0,0)$ \\
$[221]$ & $(50)$ & $(3,0)\oplus(2,1)\oplus(1,2)\oplus(1,0)\oplus(0,3)\oplus(0,1)$ \\
$[222]$ & $(36)$ & $(\frac{3}{2},\frac{3}{2})\oplus(\frac{3}{2},\frac{1}{2})\oplus(\frac{1}{2},\frac{3}{2})\oplus(\frac{1}{2},\frac{1}{2})$ \\
$[321]$ & $(45)$ & $(2,1)\oplus(1,2)\oplus(1,1)\oplus(1,0)\oplus(0,1)$ \\
$[322]$ & $(50)$ & $(3,0)\oplus(2,1)\oplus(1,2)\oplus(1,0)\oplus(0,3)\oplus(0,1)$ \\
$[331]$ & $(60)$ & $(\frac{5}{2},\frac{1}{2})\oplus(\frac{3}{2},\frac{3}{2})\oplus(\frac{3}{2},\frac{1}{2})\oplus(\frac{1}{2},\frac{5}{2})\oplus(\frac{1}{2},\frac{3}{2})\oplus(\frac{1}{2},\frac{1}{2})$ \\
$[332]$ & $(60)$ & $(\frac{5}{2},\frac{1}{2})\oplus(\frac{3}{2},\frac{3}{2})\oplus(\frac{3}{2},\frac{1}{2})\oplus(\frac{1}{2},\frac{5}{2})\oplus(\frac{1}{2},\frac{3}{2})\oplus(\frac{1}{2},\frac{1}{2})$ \\
$[333]$ & $(64)$ & $(2,1)\oplus(2,0)\oplus(1,2)\oplus 2(1,1)\oplus(1,0)\oplus(0,2)\oplus(0,1)$ \\
\hline
\hline
\end{tabular}
\end{table}

\subsubsection{\label{sec:color-magnetic}Color magnetic interaction}
The one-gluon-exchange interaction between quarks is particularly important in estimating the properties of multibaryon states. According to the quark wave function in formula (\ref{EQ:RMF}), the color current of quark $i$ can be expressed as follows:
\begin{eqnarray}
 J_i^{0C}(\vec{r_i}) &=& \frac{\lambda_i^C}{2}\frac{1}{r_i^2}I_i(r_i)|Y_{jm}^l(\theta,\phi)|^2, \label{eq:color-current1} \\
\vec{J}_i^C(\vec{r}_i) &=& \frac{\lambda_i^C}{2}\frac{1}{r_i^2}M_i(r_i)|Y_{jm}^l(\theta,\phi)|^2\left( \frac{\vec{r}_i}{r_i}\times\vec{\sigma_i} \right), \label{eq:color-current2}
\end{eqnarray}
where
\begin{eqnarray}
    I_i(r_i) &=& |G_i(r_i)|^2+|F_i(r_i)|^2,    \label{eq:Ii(ri)} \\  
  M_i(r_i) &=& G_i^*(r_i)F_i(r_i) + G_i(r_i)F_i^*(r_i).  \label{eq:Mi(ri)}
\end{eqnarray}
In this work, we consider the cases where  $u$, $d$, and $s$ quarks occupy only the first orbital $1s_{1/2}$ (with $l =m= 0$), therefore  $|Y_{jm}^l(\theta,\phi)|^2 = 1/4\pi$. The contribution from one-gluon-exchange interaction can be regarded as a correction to the mass formula (\ref{eq:actual-M}), i.e.,
\begin{align}
\Delta M &= \sum_{i<j} \alpha_s^{ij} \iint \frac{d^3 r_i \, d^3 r_j}{|\vec{r}_i - \vec{r}_j|} \left\langle J^{\mu}_{i}(\vec{r}_i) J^{C}_{\mu,j}(\vec{r}_j) \right\rangle \nonumber \\
&= \Delta M^e + \Delta M^{m},
\label{eq:deltaMG}
\end{align}
which is divided into the color-electric $\Delta M^e$ and color-magnetic parts $\Delta M^m$. In this work, to simplify our calculation, we neglect the self-energy diagram and the color-electric part of one-gluon-exchange interaction $\Delta M^e$, which are expected to be addressed in the quark mass scaling by properly readjust the parameters as they are spin-independent~\cite{Chen2012_CPC36-947}. Then Eq.~(\ref{eq:deltaMG}) for multibaryon states can be reduced into
\begin{equation}
\label{eq:c-m-mass}
\Delta M = \sum_{i<j} \frac{2\alpha_s^{ij}}{3}a_{i,j}\int_0^R M_i(r_i)\int_0^{r_i}\frac{M_j(r)}{r}\mathrm{d}r\mathrm{d}r_i.
\end{equation}
Here $\alpha_s^{ij}$ are the strong coupling constants for the one-gluon-exchange interaction between quark $i$ and quark $j$, which are in principle running with the energy scale along with the parameters $f$ and $C$~\cite{Vermaseren1997_PLB405-327, Peng2006_PLB634-413}. In this work, nevertheless, we treat them as constants and adjust them according to the mass spectra of baryons and dibaryon $D_{03}$. The interaction coefficients $a_{i,j}$ are fixed according to the classification of $N$-quark states illustrated in Sec.~\ref{sec:quark classification}. As illustrated in Ref.~\cite{Aerts1978_PRD17-260}, the coefficient $a_{i,j}$ can be fixed as follows
\begin{eqnarray}  
a_{ns} &=& N(N-10) - \frac{3}{4}N_n^2 - \frac{3}{2}N_s^2 + N_n + N_s \nonumber\\
&&{} + 4C_3 + C_4 - 4\vec{I}^2 + \frac{4}{3}\vec{J}^2 - \frac{4}{3}\vec{J}_n^2 + \frac{2}{3}\vec{J}_s^2, \nonumber\\
a_{nn} &=& \frac{3}{4}N_n^2 - N_n - C_4 + 4\vec{I}^2 + \frac{4}{3}\vec{J}_n^2, \nonumber\\
a_{ss} &=& \frac{3}{2}N_s^2 - N_s - \frac{2}{3}\vec{J}_s^2,
\end{eqnarray}
where $N_n$ and $N_s$ denote the nonstrange and strange quark numbers with $N=N_n+N_s$, respectively. $C_3$ and $C_4$ are the second-order Casimir operators' expectations for the SU(3) and SU(4) groups, which can be fixed based on Tables~\ref{tab:su(3)-su(2)} and \ref{tab:Y-SU(4)-Js)}. The irreps of SU(3), after decomposition as shown in Table~\ref{tab:su(3)-su(2)}, are given by $[\nu_1,\nu_2,\nu_3]$. By defining $p = \nu_2 - \nu_3$ and $q = \nu_3 - \nu_1$, the quadratic Casimir operator $C_3$ can be expressed as 
\begin{equation}
C_3 = \frac{1}{3}(p^2 + pq + q^2 + 3p + 3q).
\end{equation}
Similarly, the quartic Casimir operator $C_4$ for SU(4) is given by 
\begin{equation}
C_4 = 2N_s(N_s + 1) + \frac{1}{4}N(N - 18) - \frac{4}{3}(N - 9)Y.
\end{equation}
Here $Y = \frac{1}{3}N - N_s$ represents the hypercharge, $\vec{I}$ the isospin, $\vec{J_n}$ the total spin of the nonstrange quarks, $\vec{J}_s$ the total spin of the strange quarks, and $\vec{J}$ the total spin of the entire system. When the quantum numbers corresponding to the system are given, the contribution to the energy correction of the color-magnetic part of the one-gluon exchange interaction can be calculated, and the total energy can be obtained:
\begin{equation}
    \label{eq:M-oge-cmc}
    M = M_0+\Delta M.
\end{equation}

\subsection{\label{sec:Bayesian}Bayesian inference approach}
The core concept of Bayesian parameter inference is the Bayesian theorem, which is expressed as follows:
\begin{equation}
    \label{eq:bays}
    P(\mathcal{M}|D) = \frac{P(D|\mathcal{M})P(\mathcal{M})}{\int P(D|\mathcal{M})P(\mathcal{M})\mathrm{d}\mathcal{M}},
\end{equation}
Within this framework, $P(\mathcal{M}|D)$ represents the posterior probability of model $\mathcal{M}$ given the dataset $D$, which is the core objective of our analysis. $P(D|\mathcal{M})$ is the likelihood function, representing the conditional probability that model $\mathcal{M}$ correctly predicts the data $D$. $P(\mathcal{M})$ is the prior probability of model $\mathcal{M}$ before considering the data. The denominator of Eq.~(\ref{eq:bays}) serves as a normalization constant and is used to compare multiple models to determine which one can better explain the experimental data.

In this work, there are six model parameters, i.e., $p_{i = 1, 2, ..., 6} = C, \sqrt{D}, f, \alpha_s^{nn}, \alpha_s^{ns}, \alpha_s^{ss}$. By providing the quantum numbers of the multibaryon state and the appropriate model parameters, we can fix various physical properties of baryons and multibaryon states. In our previous studies~\cite{You2022_NPR39-302}, we used the least squares method to fit the model parameters and found they follow the following functional relations, i.e.,
\begin{eqnarray}   
    \sqrt{D}/\mathrm{MeV} &=& 90\times1.443^{-C}+60.5,  \label{eq:params-relation} \\
    f &=& 1.27^{-C}+1.4,  \label{eq:f-C}\\
    \alpha_s^{nn} &=& 0.97\times 2.14^C +0.2, \label{eq:params-relation3} \\
    \alpha_s^{ns} &=& 0.79\times 2.14^C +0.09, \label{eq:params-relation4} \\  
    \alpha_s^{ss} &=& 0.2.  \label{eq:params-relation5}
\end{eqnarray}
To better conduct Bayesian inference, we introduce a small offset to these formulae to generate new parameters for the model and conduct uniform random sampling within its prior range (see Table~\ref{tab:parameter_ranges}).
\begin{table}[ht]
\centering
\caption{Prior ranges for the offsets to the parameters obtained with Eqs.~(\ref{eq:params-relation}-\ref{eq:params-relation5}).}
\label{tab:parameter_ranges}
\begin{tabular}{lcc}
\hline
\hline
Parameters   & Lower limit & Upper limit \\ \hline
\midrule
$C$    & $-1.0$    & $0.5$    \\
$\delta \sqrt{D}$/MeV    & $-5.0$    & $5.0$    \\
$\delta f$    & $-0.05$    & $0.15$    \\
$\delta \alpha_s^{nn}$    & $-0.1$    & $0.1$    \\
$\delta \alpha_s^{ns}$    & $-0.1$    & $0.15$    \\
$\delta \alpha_s^{ss}$    & $-0.1$    & $0.2$    \\
\hline
\hline
\end{tabular}
\end{table}

Once the six parameters obtained through sampling are substituted into the model, the resulting theoretical mass $m_{\mathrm{th},j}$ is then used to assess the likelihood of the selected parameters reproducing the experimental mass $m_{\mathrm{exp},j}$ , where $j$ ranges from 1 to 9, as presented in the dataset $D(m_{1,2,...9})$ in Table~\ref{tab:exp-data}.

\begin{table}[ht]
\centering
\caption{Data for the masses $m_{\mathrm{exp}}$ of baryons and $D_{03}$ used in the present work~\cite{PDG2020_PTEP2020-083C01}. The most probable values of the eight baryons and $D_{03}$ along with their corresponding credible intervals (68$\%$ and 90$\%$) predicted by our model $m_{\mathrm{th}}$ are indicated as well. }
\label{tab:exp-data}
\begin{tabular}{l|cc|ccc}
    \hline  \hline
     & $m_{\mathrm{exp}}$ & $\sigma$ & $m_{\mathrm{th}}($68$\%)$ & $m_{\mathrm{th}}($90$\%)$  \\   
     & MeV  & MeV & MeV & MeV \\ \hline  
        $p$           & 938.272046  & 3.0   & $937.7_{-2.1}^{+3.5} $ & $937.7_{-3.7}^{+5.4}$     \\
        $\Lambda$     & 1115.6836   & 3.0   & $1114.7_{-1.6}^{+4.1}$ & $1114.7_{-3.4}^{+5.8}$     \\
        $\Sigma^+$    & 1189.377    & 3.0   & $1187.3_{-2.2}^{+3.1}$ & $1187.3_{-3.8}^{+4.8}$     \\
        $\Delta^{++}$ & 1232        & 3.0   & $1236.0_{-2.6}^{+3.1} $& $1236.0_{-4.3}^{+4.9}$ \\
        $\Xi^0$       & 1314.86     & 3.0   & $1317.7_{-3.6}^{+2.8}$ & $1317.7_{-5.4}^{+4.7} $    \\
        $\Sigma^{*+}$ & 1382.8      & 10.0  & $1391.2_{-3.8}^{+5.0}$ & $1391.2_{-6.6}^{+7.6}$     \\
        $\Xi^{*0}$    & 1531.8      & 10.0  & $1532.9_{-7.0}^{+5.0}$ & $1532.9_{-10.8}^{+8.4}$    \\
        $\Omega^-$    & 1672.4529   & 10.0  & $1660.6_{-12.2}^{+2.6}$& $1660.6_{-16.5}^{+7.2}$    \\
        $D_{03}$      & 2357        & 10.0  & $2333.3_{-4.3}^{+6.1} $& $2333.3_{-7.6}^{+9.3}$     \\ \hline  \hline      
    \end{tabular}
\end{table}

Then the likelihood is obtained with:
\begin{equation}
    \label{eq:likelihood}
    P_m[D(m_{1,2...9}|\mathcal{M}(p_{1,2...6})] = \underset{j = 1}{\overset{9}{\prod}}\frac{\mathrm{exp}\left[ -\frac{(m_{\mathrm{th},j}-m_{\mathrm{exp},j})^2}{2\sigma_j^2} \right]}{\sqrt{2\pi}\sigma_j},
\end{equation}
The uncertainties $\sigma_j$ on the baryon masses are negligible, while in this work we are interested in examine the possible stable multibaryons with small masses and fix the associated uncertainties in model predictions. In such cases, as indicated in Table~\ref{tab:exp-data}, we set the uncertainties  $\sigma_j=3$ MeV for baryons with masses less than 1.35 GeV and $\sigma_j=10$ MeV for the rest, which are expected to give better predictions on the masses of light multibaryons  as the parameters $C, f, \alpha_s^{nn}, \alpha_s^{ns}, \alpha_s^{ss}$ are better constrained with the energy scale close to that of light baryons.

A Markov-Chain Monte Carlo (MCMC) method using the Metropolis-Hastings algorithm is employed to simulate the posterior probability density function (PDF)
of the model parameters. The PDFs of individual parameters and their pairwise correlations are derived through marginalization over the remaining parameter space. Specifically, the marginal PDF for the $i_{th}$ parameter $p_i$ is given by:
\begin{equation}
    \label{eq£»pdf}
    P(p_i|D) =\frac{\int P(D|\mathcal{M})dp_1dp_2...dp_6}{\int P(D|\mathcal{M})P(\mathcal{M})dp_1dp_2...dp_6}.
\end{equation}
Following the standard MCMC algorithm, we discard the initial burn-in samples (the first 50,000 steps) as the chain has not yet converged to the target equilibrium distribution. The subsequent samples are retained for posterior inference, enabling robust estimation of the six-dimensional parameter probability density functions.

\section{\label{sec:Results} Results and discussions}
Employing the equivparticle model described in Sec.~\ref{sec:equivparticle} with MFA illustrated in Sec.~\ref{Sec:MFA}, we can obtain the mean-field contribution for the energy of any baryon or multibaryon state at a fixed baryon number $B$ via iteration, where the center-of-mass correction outlined in Sec.~\ref{sec:c-mass-c} is applied. With fixed quantum numbers according to the symmetry decomposition indicated in Sec.~\ref{sec:color-magnetic}, the energy correction from the color-magnetic interaction can be fixed by Eq.~(\ref{eq:c-m-mass}), then the mass of any baryon or multibaryon is obtained with Eq.~(\ref{eq:M-oge-cmc}). 

\begin{table}[ht]
\centering
\caption{Values for the most probable parameter set along with the corresponding single-parameter posterior confidence interval (68$\%$ and 90$\%$). The prior distribution is shown in Table~\ref{tab:parameter_ranges}.}
\label{tab:posterior-params}
\begin{tabular}{lcc}
\hline
\hline
Parameters   & 90$\%$ & 68$\%$ \\ \hline
\midrule
$C$    & $0.159_{-0.289}^{+0.302}$    & $0.159_{-0.100}^{+0.251}$    \\
$\delta \sqrt{D}$/{MeV}    & $0.051_{-0.328}^{+0.501}$    & $0.051_{-0.176}^{+0.299}$    \\
$\delta f$    & $0.064_{-0.047}^{+0.056}$    & $0.064_{-0.027}^{+0.036}$    \\
$\delta \alpha_s^{nn}$    & $0.011_{-0.045}^{+0.034}$    & $0.011_{-0.022}^{+0.021}$    \\
$\delta \alpha_s^{ns}$    & $0.079_{-0.049}^{+0.060}$    & $0.079_{-0.027}^{+0.045}$    \\
$\delta \alpha_s^{ss}$    & $0.149_{-0.185}^{+0.037}$    & $0.149_{-0.137}^{+0.011}$    \\
\hline
\hline
\end{tabular}
\end{table}

\begin{figure*}
\includegraphics[width=\linewidth]{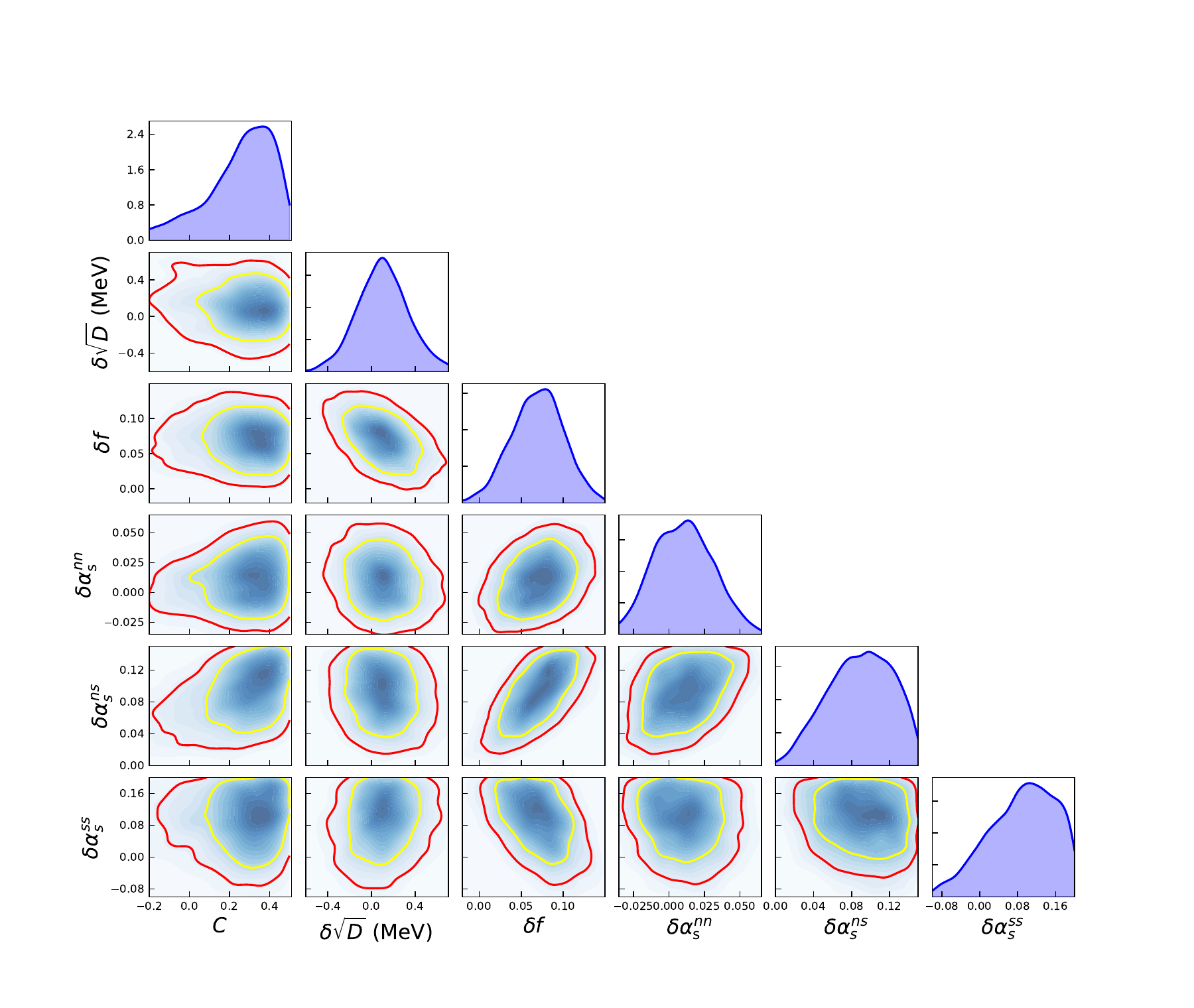}
\caption{\label{Fig:para} The PDFs of the six parameters (represented by the blue curve) and their correlations (represented by the red and yellow curves, corresponding to the 68$\%$ and 90$\%$ confidence intervals) obtained through Bayesian analysis from the masses of baryons and $D_{03}$ listed in Table~\ref{tab:exp-data}.}
\end{figure*}

To calibrate the model parameters and achieve a reasonable uncertainty estimation of our model predictions, the Bayesian parameter inference described in Sec.~\ref{sec:Bayesian} is adopted, where the experimental hadron masses illustrated in Table~\ref{tab:exp-data} are used. As specified in Table~\ref{tab:parameter_ranges}, the prior distributions of the parameters are assumed to be uniform. The obtained constraints on parameters are presented in Fig.~\ref{Fig:para}, which shows the PDFs of the six parameters from Bayesian analysis, as well as their 90$\%$ and 68$\%$ posterior credible intervals.  The probability distributions for the parameters obtained here basically exhibit a single peak structure, which means that the experimental masses of baryons and $D_{03}$ used in this paper have relatively good constraints on the six parameters of the model. The corresponding most probable values and the upper and lower limits of 68$\%$ and 90$\%$ credible intervals are indicated in Table~\ref{tab:posterior-params}. Here, parameter $C$ is relatively free and can be sampled from $-1$ to 0.5 as indicated in Table~\ref{tab:parameter_ranges}. The other parameters are sampled and combined with those estimated with functions (\ref{eq:params-relation})-(\ref{eq:params-relation5}) at fixed $C$. Note that the color-electric part of the one-gluon-exchange interaction is neglected in this work, whose effects are nonetheless expected to be included in the mass scaling in Eq.~(\ref{eq: mass scaling final}). In particular, $C > 0$ corresponds to the first-order perturbative interaction~\cite{Chen2012_CPC36-947}, while at $C < 0$ the one-gluon-exchange interaction dominates~\cite{Xia2014_PRD89-105027}. The peak value of $C$ lies in between 0.2 and 0.4, which becomes 0.159 in the most probable parameter set. This indicates a repulsive interaction among quarks and could be attributed to the first-order perturbative interaction. It is worth mentioning that the peak of the posterior marginal distribution of $\delta \sqrt{D}$ is almost at zero, which means that the Eq.~($\ref{eq:params-relation}$) accurately describes the evolution of $\sqrt{D}$ with respect to $C$. The peak in the PDF of $\delta f$ is located between 0.05 and 0.1, which means that the running parameter $f$ of the current mass of quarks is larger than that of Eq.~($\ref{eq:f-C}$). The broader peaks observed for the PDFs of $\alpha_s^{ns}$ and $\alpha_s^{ss}$ than that of  $\alpha_s^{nn}$ suggest larger parameter uncertainties, which stem from the fact that the uncertainties $\sigma$ for light baryons are smaller than that of heavy baryons as indicated in Table~\ref{tab:exp-data}.  From the perspective of the peak of the posterior distribution, the strong coupling constants seem to be underestimated in Eqs.~(\ref{eq:params-relation3}-\ref{eq:params-relation5}), and its peak are all positive, indicating  larger contributions in the energy correction of color-magnetic part of the one-gluon-exchange interaction.

Based on the two-dimensional joint posterior PDFs in Fig.~\ref{Fig:para}, the correlations between parameters can be identified. For example, there is a relatively obvious positive linear correlation between $\delta f$ and $\delta \alpha_s^{ns}$, which is mainly attributed to the similar role in varying strange quark masses and strange-nonstrange quark interactions. Slight positive correlations between the parameter pairs  $\delta f$-$\delta \alpha_s^{nn}$, $\delta \alpha_s^{ns}$-$\delta \alpha_s^{nn}$, and negative correlations $\delta f$-$\delta \alpha_s^{ss}$, $\delta f$-$\delta \sqrt{D}$ are observed. For the rest parameter pairs, the correlations are not distinctive with  the two-dimensional joint posterior PDFs showing almost circular shape.

\begin{figure*}
\includegraphics[width=\linewidth]{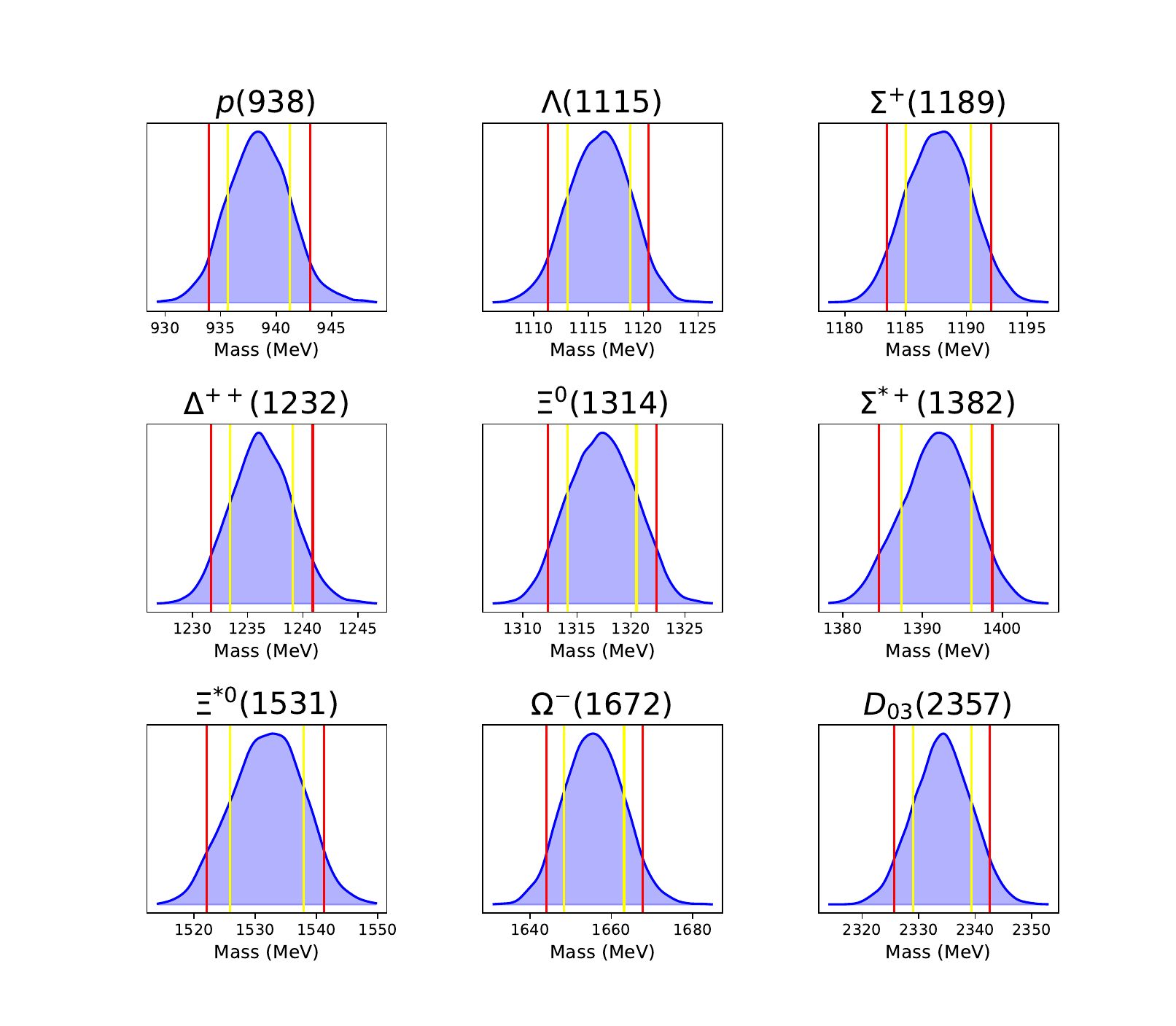}
\caption{\label{Fig:posterior-baryon} The PDFs for the masses (in MeV) of the 8 baryons and $D_{03}$ obtained adopting the prior ranges of  parameters listed in Table~\ref{tab:parameter_ranges} and the constraints in Table~\ref{tab:exp-data}, where the red (yellow) vertical lines represent the 90$\%$ (68$\%$) credible interval.}
\end{figure*}


The obtained PDFs for the masses of the 8 baryons and $D_{03}$ employed for Bayesian inference are presented in Fig.~\ref{Fig:posterior-baryon}, while the most probable values along with their 68$\%$ and 90$\%$ credible intervals are listed in Table~\ref{tab:exp-data}. Note that the most probable values listed in Table~\ref{tab:exp-data} are fixed adopting the most probable parameter set instead of the median values for each hadron. Evidently, for light baryons such as nucleons $N$, $\Lambda$, $\Sigma$,  $\Xi$, and $\Delta$, our model generally well reproduces their experimental masses, which are enclosed within the 68$\%$ intervals with $\sigma\approx 3$ MeV. For heavier baryons, the deviation from the experimental masses grows quickly and approaches to $\sigma\approx 10$ MeV. For example,  the most probable mass of $\Omega^-$ predicted by the model differs from the experimental value by approximately 12 MeV and does not fall within the 90$\%$ credible interval. For the spin partner states $\Sigma^+$ and $\Sigma^{*+}$, the most probable mass of $\Sigma^{+}$ with $J = 1/2$ is very close to the experimental value and lies within the 68$\%$ confidence interval, while that of $\Sigma^{*+}$ is approximately 9 MeV larger than the experimental value and exceeds the 90$\%$ confidence interval. For $\Delta^{++}$ resonant state, although the experimental value differs from its most probable mass by 4 MeV, it still falls within the 90$\%$ credible interval. In general, the credible intervals are roughly consistent with the uncertainties $\sigma_i$ adopted for Bayesian inference and enclose the experimental masses, while model predictions on the mass spectra are consistent with the experimental values for light baryons with masses below 1.35 GeV.

Beside baryons, the CELSIUS/WASA collaboration observed a clear enhancement in the $\pi\pi$ and $d\pi$ invariant mass spectra in the $pn \rightarrow d\pi^0\pi^0$ reaction at 1.03 and 1.35 GeV, suggesting an s-channel resonance (dibaryon $D_{03}$ with mass 2.36 GeV and width 80 MeV), as t-channel $\Delta \Delta$ processes alone cannot explain the effect~\cite{Bashkanov2009_PRL102-052301}. Therefore, we adopt the experimental mass 2357 MeV for $D_{03}$ to constrain the parameters as indicated in Table~\ref{tab:exp-data}~\cite{Wang2017_NPR34-001, Huang2014_PRC89-034001}. Note that partial wave analysis revealed a pole at $2380 \pm 10 - {i}(40 \pm 5)$ MeV in the NN $ ^3 D_3 - ^3 G_3$ channel, suggesting a larger mass for the dibaryon $D_{03}$~\cite{Adlarson2016_PLB762-455}. In this work, the most probable mass of $D_{03}$ predicted by our model is 2333.3 MeV, which is smaller than the experimental values. Nevertheless, our calculation is generally consistent with the estimation of  a constituent quark model predicting 2342 MeV for the mass of $D_{03}$~\cite{Kim2020_PRD102-074023}, which falls within the 90$\%$ credible interval in our current estimation. The most probable mass relative $\Delta \Delta(2464)$ threshold binding energy is 131 MeV, while the HAL QCD collaboration group recently provided a binding energy of approximately 30 to 40 MeV for the double $\Delta$ system with the pion mass being about 2 to 3 times its actual value~\cite{Gongyo2020_PLB811-135935}. Note that in this work we treat the $D_{03}$ state as a compact hexaquark state instead of a molecular state, which could be the reason for the large differences in the binding energies.

\begin{figure}
\includegraphics[width=0.9\linewidth]{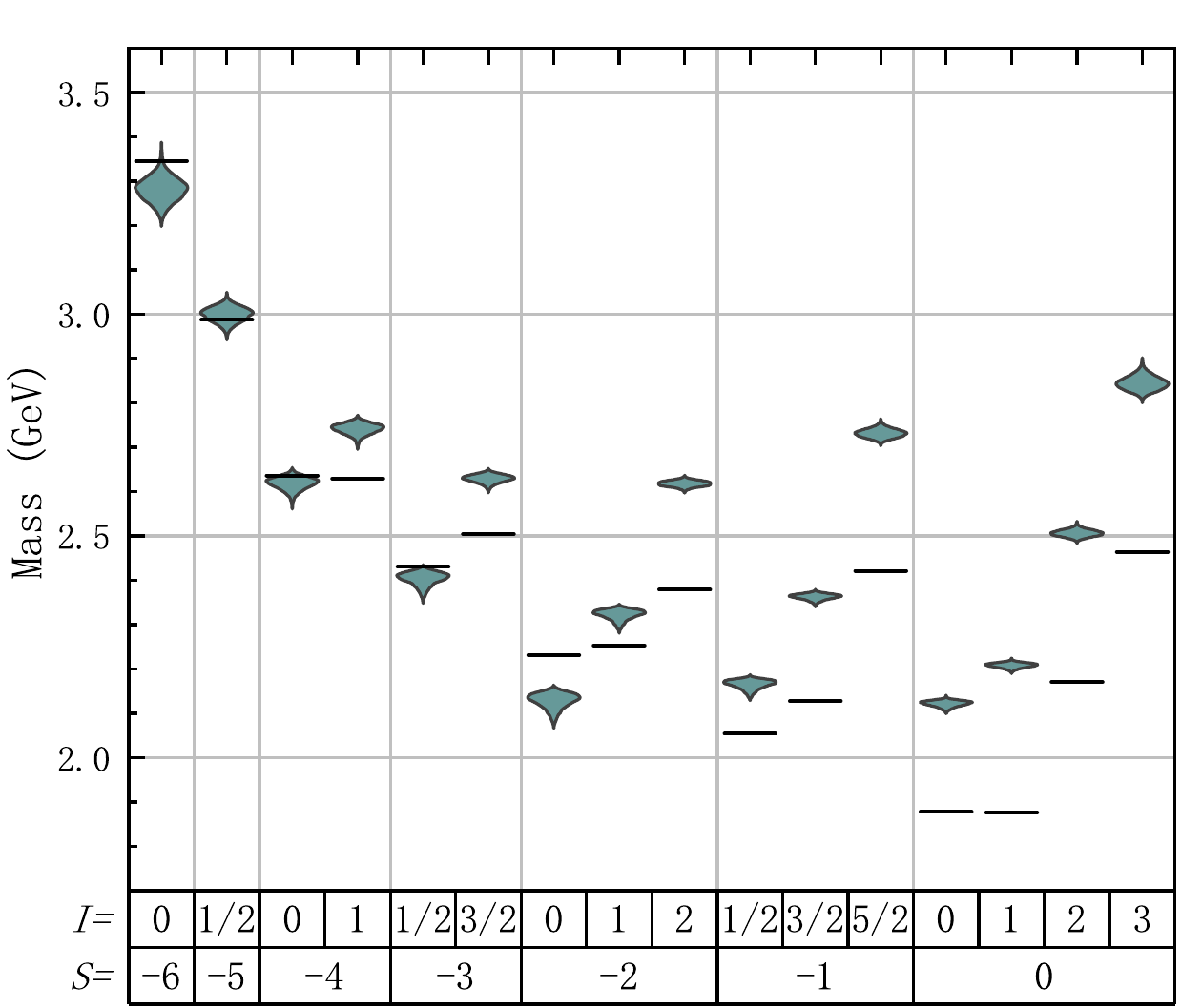}
\caption{\label{Fig:A2}The posterior mass distribution for most stable compact dibaryons ($B=2$) at fixed strangeness $S$ and isospin $I$, which are fixed according to the PDFs of the parameters presented in Fig.~\ref{Fig:para}. The dash lines indicate the thresholds for the minimum masses of two baryons with the same quark contents.}
\end{figure}

\begin{figure}
\includegraphics[width=0.9\linewidth]{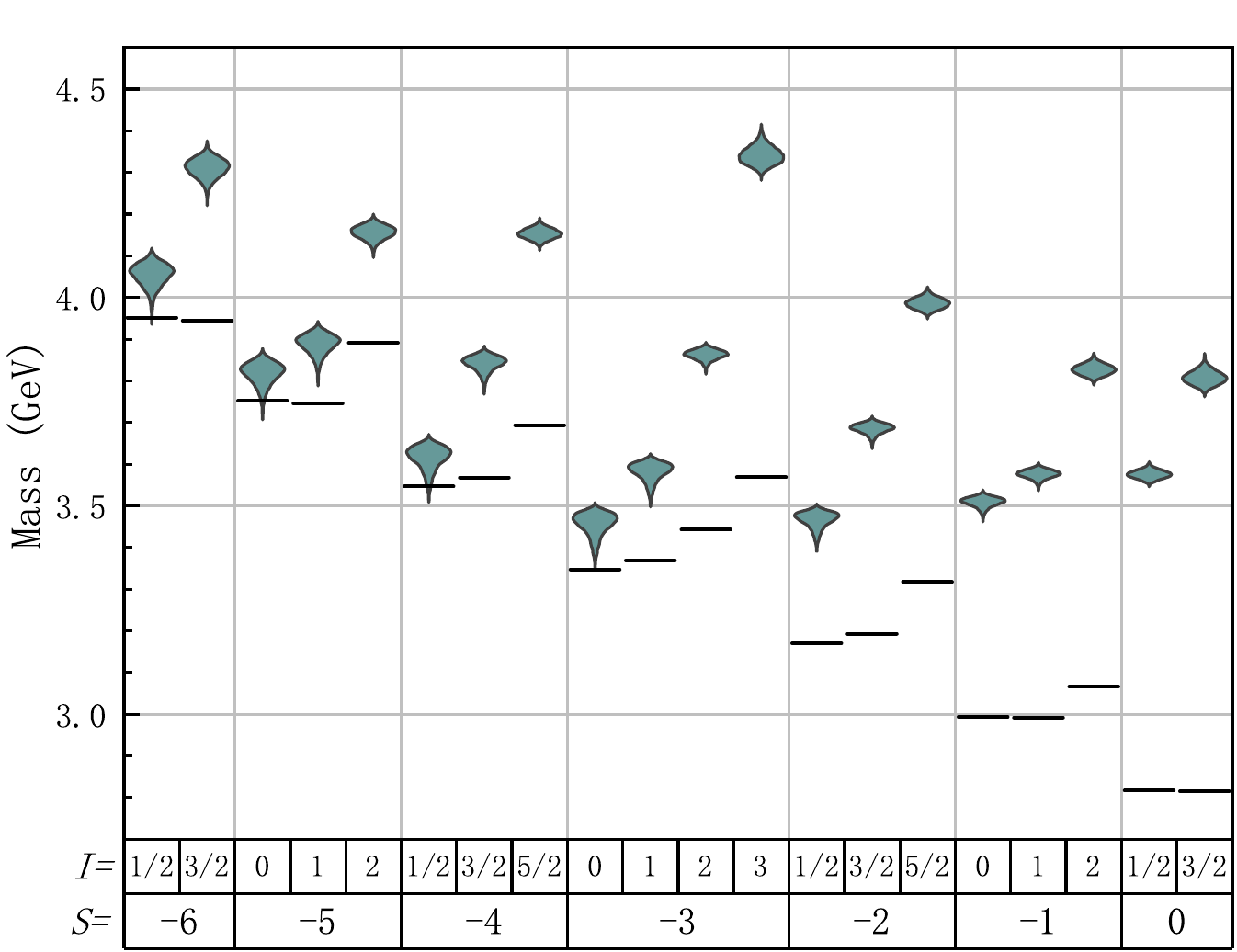}
\caption{\label{Fig:A3} Same as Fig.~\ref{Fig:A2} but for compact tribaryons with baryon number $B=3$.}
\end{figure}

\begin{figure}
\includegraphics[width=0.9\linewidth]{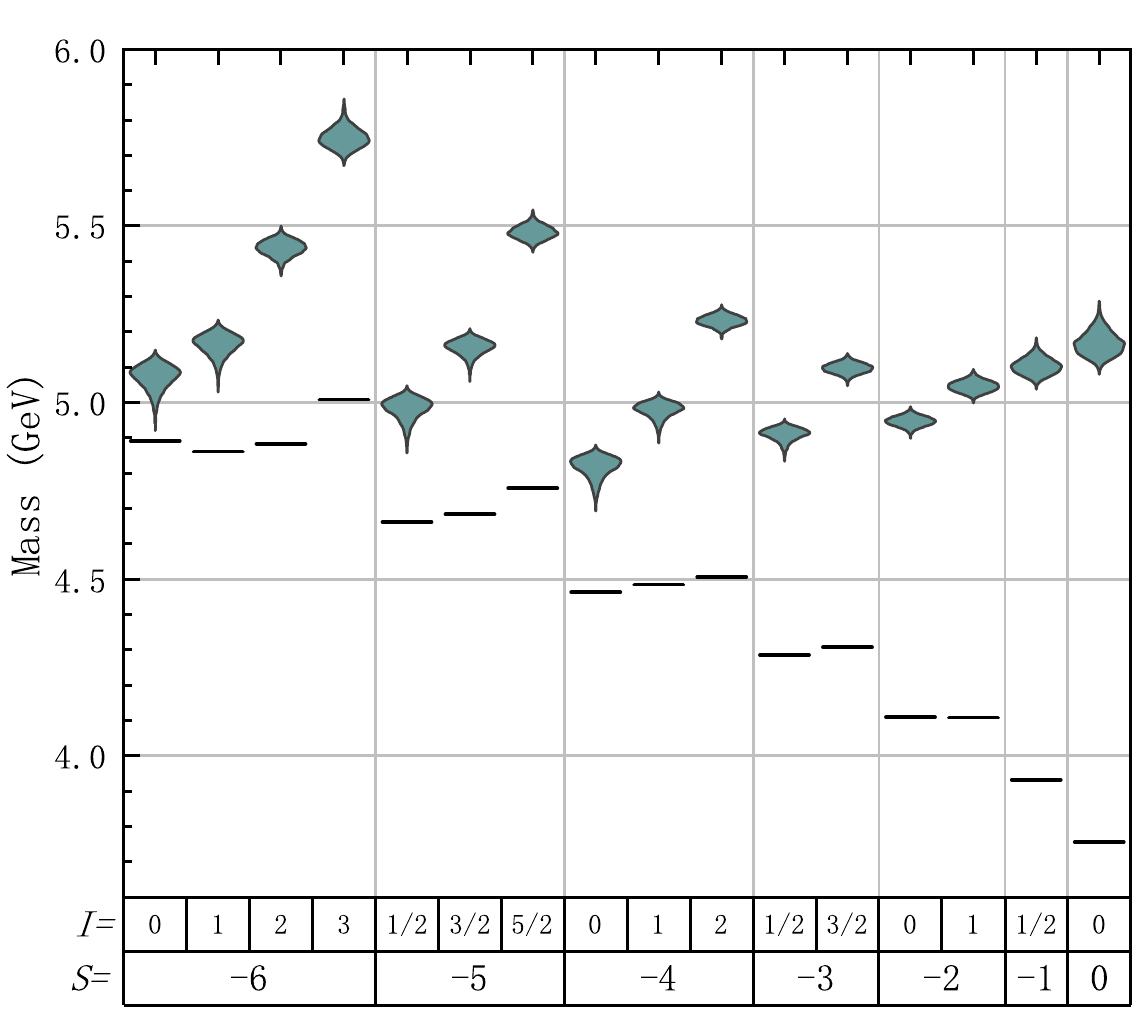}
\caption{\label{Fig:A4} Same as Fig.~\ref{Fig:A2} but for compact multibaryon states with baryon number $B=4$.}
\end{figure}

\begin{figure}
\includegraphics[width=0.9\linewidth]{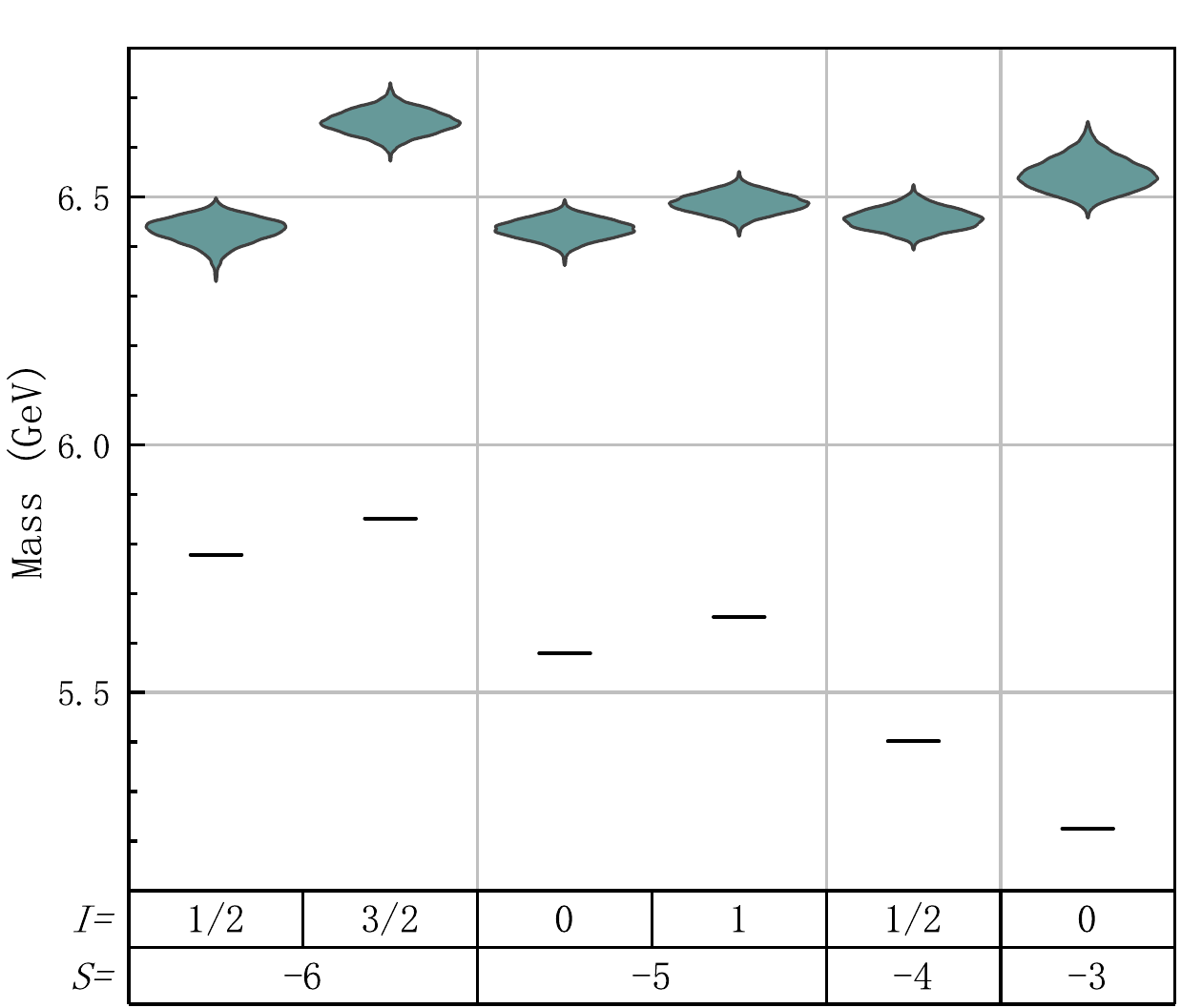}
\caption{\label{Fig:A5} Same as Fig.~\ref{Fig:A2} but for compact multibaryon states with baryon number $B=5$.}
\end{figure}

\begin{figure}
\includegraphics[width=0.9\linewidth]{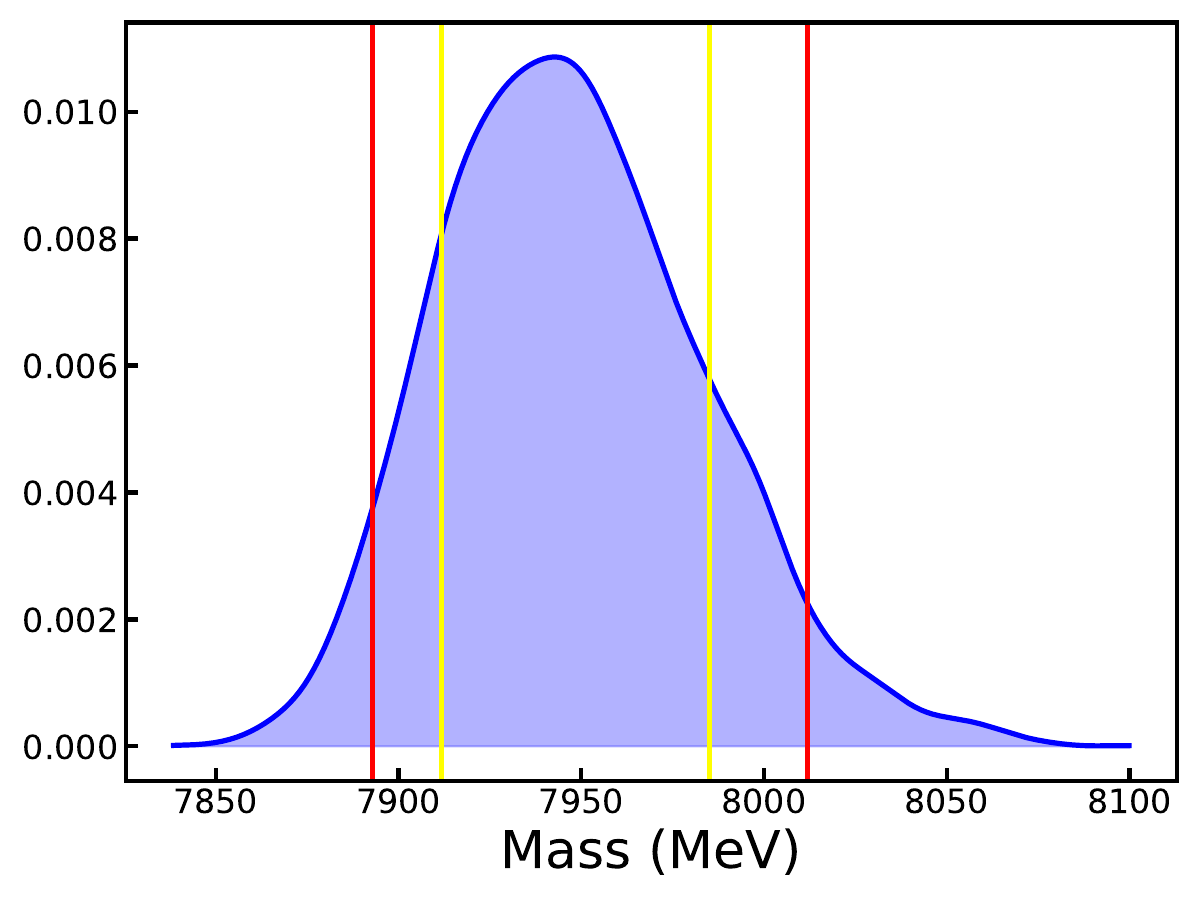}
\caption{\label{Fig:A6} The PDF for the mass (in MeV) of the multibaryon state ($B=6$) fixed according to the posterior distribution of the parameters in Table~\ref{tab:posterior-params}, where the red (yellow) vertical lines represent the 90$\%$ (68$\%$) credible interval.}
\end{figure}

Based on the posterior distribution of the parameters for the equivparticle model, we then conduct a systematic study on the mass spectra of multibaryon states with baryon numbers $B$ ranging from 2 to 6, where the complete results are listed in the supplementary file. Since the masses of multibaryon states vary little with respect to the third component of isospin $I_3$ due to the slight variations in Coulomb energy, in this work we consider only the cases with $I_3 = I$ for simplicity. The obtained results for various multibaryon states with $B=2$-6 are then presented in Figs.~\ref{Fig:A2}-\ref{Fig:A6}, where the lowest energy state at fixed baryon number $B$, isospin $I$, and strangeness $S$ are indicated.

In Fig.~\ref{Fig:A2}, we present the posterior mass distribution for the most stable compact dibaryons at fixed strangeness $S$ and isospin $I$  with baryon number $B = 2$. The dashed lines indicate the lowest mass threshold for the two baryons in vacuum with the same quark composition. There are possibilities that several stable dibaryons with the masses smaller than the threshold of two baryons in vacuum. 
For example, the six-quark state with quantum numbers $S = -6, I = 0$, and $ J^P = 0^+$ is stable according to the $\Omega\Omega$ threshold indicated by the thin horizontal line. In our current study, the most probable mass of this state is 3307.621 MeV, with a 90$\%$ credible interval of $3307.621_{-70.726}^{+18.735}$ MeV and a 68$\%$ credible interval of $3307.621_{-53.00}^{+0.62}$ MeV. In particular, there is a 98.44$\%$ probability that the mass of this dibaryon is below the mass threshold of $\Omega\Omega$, i.e., it is likely a bound state with the most probable binding energy being 37.25 MeV. In literature, this dibaryon was also predicted by Kopeliovich in the framework of the flavor SU(3) Skyrmion model~\cite{Kopeliovich1998_NPA639-75c}. Both ChQM and the QDCSM model have predicted the dibaryon to be bound~\cite{Pang2002_PRC66-025201, Huang2020_PRC101-034004}, where the obtained binding energies lie within the 90$\%$ credible interval of the posterior mass in our current study. The HAL QCD Collaboration employing (2+1)-flavor lattice QCD simulations with near-physical pion mass ($m_{\pi} \approx 146$ MeV) has investigated this state as well, and demonstrating that the $\Omega\Omega$ system exhibits net attractive interaction with a binding energy of approximately 1.6 MeV~\cite{Gongyo2018_PRL120-212001}. 

For the H-dibaryon with quantum numbers $S=-2$, $I = 0$, $J = 0$ and the dimensionality of the SU(3) irrep being 1, as indicated in Fig.~\ref{Fig:A2}, the most probable mass and the 90$\%$ credible interval are $2120.691_{-25.919}^{+31.403}$ MeV, indicating a binding energy of $109.309_{-31.403}^{+25.918}$ MeV with respect to the $\Lambda\Lambda$ system. This state is consistent with previous investigations and was first proposed in 1977 as a deeply bound 6-quark state~\cite{Jaffe1977_PRL38-195,Jaffe1977_PRL38-617}. With SU(3)-symmetric quark masses ($m_{\pi} \sim 806$ MeV) and multiple lattice volumes ($3.4 - 6.7$ fm), the NPLQCD collaboration examined the H-dibaryon and find its binding energy $B_{\mathrm{H}} = 74.6 \pm 3.3$ MeV~\cite{Beane2013_PRD87-034506}, which is close to the lower limit of the 90$\%$ credible interval obtained in this work. NPLQCD also reported a candidate excited state in the $\mathrm{SU}(3)$ $\mathbf{27}$-plet representation ($B \approx 15.9\,\mathrm{MeV}$)~\cite{Beane2013_PRD87-034506}. However, our prediction indicates that these $\mathbf{27}$-plet states are unbound with respect to both the $\Lambda N$ and $\Lambda \Lambda$ thresholds. HAL collaboration found a weakly bound H close to the $\Lambda\Lambda$ threshold adopting heavy $u$, $d$ quarks~\cite{Inoue2011_PRL106-162002}. According to the experimental measurements of $_{\Lambda\Lambda}^{6}$He~\cite{Ahn2013_PRC88-014003},  it is not likely that the H-dibaryon particle is deeply bound, where the effect of SU(3) flavor symmetry breaking would play an important role and should be explored in our future works.

Beside these dibaryons, the state with $I = \frac{1}{2}$ and $S = -5$ has a 23.64$\%$ probability of having a mass lower than the mass threshold of $\Omega^- \Xi^0$, while its most probable mass is 3012.899 MeV. The mass of the state with $S = -4$ and $I = 0$ has a 92.49$\%$ probability of being lower than the $\Xi^0\Xi^-$ threshold, and its most probable mass is 2619.599 MeV. 

Figure~\ref{Fig:A3} shows the most stable compact tribaryons at fixed strangeness $S$ and isospin $I$  with baryon number $B = 3$, which generally become less bound than dibaryons. Nevertheless, there are still marginal possibilities for the tribaryons to be stable according to the equivparticle model adopted here, i.e.,
\begin{enumerate}
    \item 0.25$\%$ probability for the tribaryon (most probable mass: 4064.412 MeV) with $S = -6$, $I = 1/2$, and $J = 1/2$ to be less than the $\Xi^-\Xi^0\Xi^0$ threshold;
    \item 2.19$\%$ probability for the tribaryon (most probable mass: 3817.205 MeV) with $S = -5$, $I = 0$, and $J = 3/2$ to be less than the $\Lambda\Xi^0\Xi^-$ threshold;
    \item 2.21$\%$ probability for the tribaryon (most probable mass: 3608.067 MeV) with $S = -4$, $I = 1/2$, and $J = 1/2$ to be less than the $\Lambda\Lambda\Xi^0$ threshold. 
\end{enumerate}
For the tribaryon states with $S = -2$, $I = 1/2$ and 3/2, our model prediction suggests that they are unbound according to the $\Lambda\Lambda N$ and $\Xi N N$ thresholds. Nevertheless, in the literature it was found that the tribaryon state with $(I)J^{P} = (\frac{1}{2})\frac{3}{2}^+$ is deeply bound with the mass being 13.5 MeV smaller than the $\Xi d$ threshold~\cite{Garcilazo2016_PRC93-034001}. Since this state cannot decay into the $\Lambda \Lambda N$ channel under a pure S-wave configuration, its decay width is expected to be very small~\cite{Garcilazo2016_PRC93-034001}. For tribaryons with $S = 0$, $I = 1/2$, and $3/2$, we predict that their masses all exceed the $NNN$ threshold. Compared with the $\Delta\Delta\Delta$ threshold, nevertheless, the most probable mass and 90$\%$ credible interval for the tribaryon with $S = 0$ and $I = 1/2$ is $3576.11_{-20.59}^{+11.76}$ MeV, indicating a binding energy of approximately 120 MeV. This is in good agreement with the deepest bound state of $I = 1/2$ (with a binding energy of 84 MeV) given by Garcilazo even though in this work we have assumed the tribaryon as a compact state~\cite{Garcilazo1997_PRC56-84}.


Finally, in Figs.~\ref{Fig:A4}, \ref{Fig:A5}, and \ref{Fig:A6} we present the posterior mass distributions for the most stable compact multibaryons with baryon numbers $B = 4$, 5, and 6. Our calculation suggests that they are unbound according to the thresholds of the masses of baryons with the same quark contents. Theoretical models predict the four-body hypernucleus $T \equiv (n,n,\Lambda,\Lambda)$ may form a Borromean bound state with binding energies 1-9 MeV, which is potentially detectable through the $d + d \rightarrow T + K^+ + K^+$ reaction~\cite{Richard2015_PRC91-014003}. In Fig. \ref{Fig:A6}, the red and yellow solid lines respectively represent the 90$\%$ and 68$\%$ credible intervals for the mass of multibaryon state with $B = 6$. This state is unbound according to the 6$\Lambda$ threshold. Meanwhile, a spin-zero and charge-neutral six-baryon state ($u,d,s$ combinations) as a minimal stable ``quark-$\alpha$" particle could potentially be bound by $\sim 140$ MeV/nucleon through strangeness-enhanced interactions~\cite{Michel1988_PRL60-677}.

\section{\label{sec:summary}Summary}

In the framework of an equivparticle model, we investigate systematically the mass spectra of color-singlet $N$-quark states with $N = 3, 6, 9, 12, 15$, and 18  employing mean-field approximation. In particular, we solve the Dirac equation of quarks and consider the quarks occupying only the 1s$_{1/2}$ state, where the $N$-quark states are assumed to be spherically symmetric. By considering the center-of-mass corrections and one-gluon-exchange interactions, the masses of various baryon and multibaryon states can be fixed by properly addressing the color-magnetic part of the one-gluon-exchange interaction at fixed baryon number $B=N/3$, strangeness $S$, and isospin $I$.

In order to effectively constrain the 6 free parameters of the model, we carry out a Bayesian parameter inference based on the experimental masses of eight baryons and $D_{03}$. The posterior probability density functions and their correlations of the model parameters are examined. Based on the posterior distribution of parameters, we further make predictions on the masses of multibaryon states and provide their 68$\%$ and  90$\%$ credible intervals. Our findings are listed as following:
\begin{itemize}
    \item For dibaryons ($B=2$), we identify a particularly stable state ($S=-6$, $I=0$, $J^P=0^+$) with the most probable mass being 3307.621 MeV (binding energy 37.25 MeV) and a 98.44$\%$ probability for the mass to lie below the $\Omega\Omega$ threshold. The H-dibaryon ($S=-2, I=0$) emerges as another promising candidate with a most probable mass 2120.691 MeV and binding energy  109.309 MeV, though lattice QCD studies suggest somewhat smaller binding energies~\cite{Inoue2011_PRL106-162002}. Notably, we find a $S=-4, I=0$ dibaryon state with 92.49$\%$ probability of being bound with respect to the $\Xi^0\Xi^-$ threshold (most probable mass: 2619.599 MeV), while the $S=-5, I=1/2$ configuration shows weaker binding energy (23.64$\%$ probability below $\Omega^-\Xi^0$ threshold). 
    \item For tribaryons ($B=3$), we observe marginal binding probabilities for certain configurations: 2.19$\%$ relative to $\Lambda \Xi^0\Xi^-$ threshold for a $S=-5, I=0$ state (most probable mass 3817.205 MeV) and 2.21$\%$  relative to $\Lambda\Lambda\Xi^0$ threshold for a $S=-4, I=1/2$ state (most probable mass 3608.067 MeV). We have also identified a deeply bound $I=1/2$ state with binding energy $\sim$120 MeV relative to the $\Delta\Delta\Delta$ threshold (most probable mass 3576.11 MeV), aligning well with previous theoretical predictions~\cite{Garcilazo1997_PRC56-84}. 
    \item For multibaryon stats with $B=4$-6, our calculations generally predict unbound states relative to their respective baryon thresholds. 
\end{itemize}
The systematic mass predictions across different baryon numbers and quantum states provide useful benchmarks for future experimental searches of exotic multibaryon states, particularly highlighting the H, $D_{03}$, $\Omega\Omega$, $\Omega^- \Xi^0$, and $\Xi^0\Xi^-$ dibaryons and certain tribaryon configurations as the most promising candidates for stable, compact multi-quark systems. Our results demonstrate that strangeness plays a crucial role in enhancing stability in these systems, where the multibaryon state with $I=0$ and $S=-B$ is generally more stable than the others at a fixed $B$, indicating the preference over SU(3) flavor symmetry with $N_u=N_d=N_s$. The stable and unstable multibaryon states examined in this work may persist in compact stars, which could be helpful for us to understand the essence of pulsar-like objects.

\section{Acknowledgments}
This work was supported by the National Natural Science Foundation of China (Grant No. 12275234) and the National SKA Program of China (Grant No. 2020SKA0120300 and No. 2020SKA0120100).

\bibliography{Multibaryon}

\end{document}